\documentclass{amsart}
\usepackage{amsmath}
\usepackage{amssymb}
\usepackage{amsthm}
\usepackage{amsfonts}
\usepackage{amstext}
\usepackage{amsopn}
\usepackage{amsxtra}
\usepackage{mathrsfs}
\usepackage{graphicx}
\usepackage[latin1]{inputenc}
\usepackage{dsfont}
\usepackage{cite}
\newcommand\R{{\ensuremath {\mathbb R} }}

\renewcommand\phi{\varphi}

\newcommand{\wto}{\rightharpoonup}
\renewcommand{\to}{\rightarrow}
\newcommand{\cD}{\mathcal{D}}

\newcommand\ii{{\ensuremath {\infty}}}
\newcommand\pscal[1]{{\ensuremath{\left\langle #1 \right\rangle}}}

\DeclareMathOperator{\tr}{{\rm Tr}}
\renewcommand{\leq}{\leqslant}
\renewcommand{\geq}{\geqslant}
\renewcommand{\epsilon}{\varepsilon}

\begin{document}
 
\author[M. Lewin]{Mathieu LEWIN}
\address{CNRS \& Laboratoire de Mathématiques (UMR 8088), Université de Cergy-Pontoise, F-95000 Cergy-Pontoise, France.}
\email{Mathieu.Lewin@math.cnrs.fr}

\title[Comment on a paper by C. Argaez \& M. Melgaard]{Comment on `Solutions to quasi-relativistic multi-configurative Hartree-Fock equations in quantum chemistry', by C. Argaez \& M. Melgaard}

\maketitle

\begin{abstract}
In a recent paper published in \textit{Nonlinear Analysis: Theory, Methods \& Applications}, C. Argaez and M. Melgaard studied excited states for pseudo-relativistic multi-configuration methods. Their paper follows a previous work of mine in the non-relativistic case (\textit{Arch. Rat. Mech. Anal} \textbf{171}, 2004). The main results of the paper of C. Argaez and M. Melgaard are correct, but the proofs are both \emph{wrong} and \emph{incomplete}.  

\medskip 

\noindent{\scriptsize\copyright~2011 by the author.}
\end{abstract}

\medskip

In a recent paper~\cite{ArgMel-12} published in \textit{Nonlinear Analysis: Theory, Methods \& Applications}, C. Argaez and M. Melgaard studied the existence of excited states in a nonlinear model of quantum chemistry called \emph{multi-configuration}. Their paper extends to the pseudo-relativistic setting some results which I have obtained in the non-relativistic case in~\cite{Lewin-04a}. The main results of the paper of C. Argaez and M. Melgaard are correct, but the proofs are both \emph{wrong} and \emph{incomplete}. The purpose of this letter is to explain why.

In the paper of Argaez-Melgaard, there is an important confusion between \emph{Hartree-Fock} (HF) and \emph{multi-configuration} (MC) theories. Indeed, sections 5, 7 and 8 in~\cite{ArgMel-12} have simply been copied and pasted from a previous paper~\cite{EnsMel-09} of Enstedt-Melgaard on pseudo-relativistic Hartree-Fock equations, without being of any help in the multi-configuration case.

Let me quickly re-explain the difference between the Hartree-Fock and multi-configuration models. The aim of these two nonlinear theories is to approximate the solutions of the many-body Schrödinger equation, describing electrons in atoms and molecules. The Hamiltonian of the system is the operator
\begin{equation}
H^{\rm NR/R}:=\sum_{j=1}^N\Big(T^{\rm NR/R}_{x_j}+V(x_j)\Big)+\sum_{1\leq k<\ell\leq N}\frac{1}{|x_k-x_\ell|}
\label{eq:Hamiltonian}
\end{equation}
acting on the subspace $\bigwedge_1^N L^2(\R^3)$ of antisymmetric functions $\Psi(x_1,...,x_N)$ in $L^2((\R^3)^N)$. The function $V$ is the electrostatic potential induced by the nuclei in the system which, in the Born-Oppenheimer approximation, are treated as classical pointlike particles:
$$V(x):=-\sum_{k=1}^M\frac{z_k}{|x-R_k|}.$$
Here $z_k>0$ and $R_k\in\R^3$ are, respectively, the charges and positions of the nuclei. The total nuclear charge is $Z=\sum_{k=1}^Mz_k$.
The operator $T^{\rm NR/R}$ describes the kinetic energy of the electrons. In non-relativistic quantum mechanics
\begin{equation}
T^{\rm NR}=-\frac{\Delta}{2}.
\label{eq:T_nonrel} 
\end{equation}
Relativistic particles should be described by the Dirac operator but there is no consistent theory for interacting systems at present. It is therefore useful to test some ideas on the so-called \emph{pseudo-relativistic} operator
\begin{equation}
T^{\rm R}=\alpha^{-2}\Big(\sqrt{1-\alpha^2\Delta}-1\Big). 
\label{eq:T_rel}
\end{equation}
Here $\alpha>0$ is the (bare) fine structure constant whose inverse is the speed of light. Pseudo-relativistic many-body systems based on $T^{\rm R}$ have been considered before in several important works including~\cite{LieThi-84,LieYau-87,LieYau-88,LewSieVug-97,ZhiVug-02,Zhislin-06}. The quadratic form associated with $H^{\rm R}$ is not always bounded from below (as opposed to the case of $H^{\rm NR}$) but it is so when $\alpha Z\leq 2/\pi$, by the Hardy-Kato inequality. In this case $H^{\rm R}$ is well defined by Friedrichs' method and it has similar properties as in the non-relativistic case. In particular, when $N<Z+1$ and $\alpha Z<2/\pi$, there are infinitely many eigenvalues $\lambda_k$'s below its essential spectrum~\cite{ZhiVug-02,Zhislin-06}, corresponding to Schrödinger's equation
\begin{equation}
H^{\rm NR/R}\;\Psi_k=\lambda_k\;\Psi_k.
\label{eq:eigenvalue}
\end{equation}

For any normalized wave-function $\Psi$, it is convenient to define the one-particle density matrix $\gamma_\Psi$ by its integral kernel 
$$\gamma_\Psi(x,y)=N\int_{\R^3}dx_2\cdots\int_{\R^3}dx_N\; \overline{\Psi(x,x_2,...,x_N)}\,\Psi(y,x_2,...,x_N).$$
This is a self-adjoint operator on $L^2(\R^3)$ such that $0\leq\gamma_\Psi\leq 1$ and $\tr(\gamma_\Psi)=N$.
Any one-body observable can be expressed in terms of $\gamma_\Psi$ only. For instance, 
$$\pscal{\Psi,\Bigg(\sum_{j=1}^N\Big(T^{\rm NR/R}_{x_j}+V(x_j)\Big)\Bigg)\Psi}_{L^2((\R^3)^N)}=\tr_{L^2(\R^3)}\big(T^{\rm NR/R}+V\big)\gamma_\Psi$$
where both sides are interpreted in the sense of quadratic forms. On the other hand, the electronic Coulomb repulsion (the second sum in formula~\eqref{eq:Hamiltonian}) is a two-body operator which cannot be expressed in terms of $\gamma_\Psi$ only, except for very specific states like Hartree-Fock states.

The Hartree-Fock method~\cite{LieSim-77,Lions-87} consists in restricting the many-body energy to wave-functions which are a single \emph{Slater determinant}
$$\Psi_{\rm HF}(x_1,...,x_N)=\big(\phi_1\wedge\cdots\wedge\phi_N\big)(x_1,...,x_N):=\frac{1}{\sqrt{N!}}\,{\rm det}(\phi_i(x_j)),$$
where $\pscal{\phi_i,\phi_j}=\delta_{ij}$.
Such states are completely described by their one-particle density matrix, which is the orthogonal projection on ${\rm span}(\phi_1,...,\phi_N)$:
$$\gamma_{\Psi_{\rm HF}}=\sum_{j=1}^N|\phi_j\rangle\langle\phi_j|.$$
In particular, the two-body energy (hence also the total energy) can be expressed in terms of $\gamma_{\Psi_{\rm HF}}$ only. This fact is used in modern theoretical studies of HF-type models, as pioneered by Bach, Lieb, and Solovej~\cite{Lieb-81,Solovej-91,Bach-92}, as well as in numerical optimization techniques~\cite{CanDefKutLeBMad-03}.

The multi-configuration methods are based on the observation that Slater determinants span the whole many-body space. The many-body energy is then restricted to wave-functions which are a \emph{finite} linear combination of Slater determinants:
\begin{equation}
\Psi_{\rm MC}(x_1,...,x_N)=\sum_{1\leq i_1<\cdots <i_N\leq K}a_{i_1...i_N}\;\phi_{i_1}\wedge\cdots\wedge\phi_{i_N}. 
\label{eq:form-MC}
\end{equation}
The unknowns are the \emph{mixing coefficients} $a_{i_1...i_N}$ which must satisfy the normalization constraint $\sum|a_{i_1...i_N}|^2=1$ and the \emph{orbitals} $\phi_1,...,\phi_K$ which must be orthonormal: $\pscal{\phi_i,\phi_j}=\delta_{ij}$. When $K=N$ or when all the $a_{i_1...i_N}$'s vanish except one, we are back to the HF method. The MC energy is nothing but the many-body energy of such special states, expressed in terms of the mixing coefficients  $a_{i_1...i_N}$'s and the orbitals $\phi_1,...,\phi_K$. In general, the interaction energy cannot be expressed only in terms of the one-particle density matrix.  The MC equations consist of a system of $K$ coupled nonlinear PDE's for the $\phi_j$'s, together with a $K\choose N$-dimensional eigenvalue equation for the $a_{i_1...i_N}$'s.

The existence of minimizers for non-relativistic MC was proved in a fundamental paper of Friesecke~\cite{Friesecke-03} and, later, in a paper of mine~\cite{Lewin-04a}, with a different method based on previous works by Lions~\cite{Lions-87} and Fang-Ghoussoub~\cite{FanGho-94,Ghoussoub-93}. This technique allowed me to construct specific critical points of the MC energy, interpreted as approximate excited states in a certain sense. This is what was extended to the pseudo-relativistic case in the paper of Argaez-Melgaard.

\medskip

Except for the sections copied from~\cite{EnsMel-09}, the paper of Argaez-Melgaard follows very closely my paper~\cite{Lewin-04a}. I think it is fine to copy the literature, as long as the source is clearly mentioned, which is debatable here. The method of~\cite{Lewin-04a} works also in the pseudo-relativistic case, except for some minor steps. One difficulty is that the potential term involving $V$ is \emph{not} continuous for the $H^{1/2}(\R^3)$ weak topology:
\begin{equation}
\phi_n\wto\phi \text{ weakly in $H^{1/2}(\R^3)$}\quad \not\Longrightarrow\quad \int_{\R^3}\frac{|\phi_n(x)|^2}{|x|}\,dx\to\int_{\R^3}\frac{|\phi(x)|^2}{|x|}\,dx.
\label{eq:implication} 
\end{equation}
Said differently, the operator $(1-\Delta)^{-1/4}|x|^{-1}(1-\Delta)^{-1/4}$ is \emph{not} compact. The result corresponding to~\eqref{eq:implication} is true in the non-relativistic case in which $H^{1/2}(\R^3)$ is replaced by $H^1(\R^3)$, and this was used in~\cite{Lewin-04a} as well as in most papers dealing with non-relativistic atomic models. What is true in the pseudo-relativistic case, however, is that the \emph{total} one-body energy is weakly lower semi-continuous, that is
\begin{multline}
\phi_n\wto\phi \text{ weakly in $H^{1/2}(\R^3)$}\\ \Longrightarrow\ \liminf_{n\to\ii}\pscal{\phi_n,\big(T^{\rm R}+V\big)\phi_n}\geq \pscal{\phi,\big(T^{\rm R}+V\big)\phi}
\label{eq:wlsc} 
\end{multline}
as soon as $\alpha Z<2/\pi$. This well-known  fact was already employed in pseudo-relativistic Hartree-Fock theory by Dall'Acqua, {\O}stergaard S{\o}rensen and Stockmeyer in~\cite{DalSorSto-08}. The reason why~\eqref{eq:wlsc} is true is that $T^{\rm R}+V$ can be written as 
$T^{\rm R}+V=\big(T^{\rm R}+V\big)_+ - \big(T^{\rm R}+V\big)_-$
where $\big(T^{\rm R}+V\big)_+\geq0$ and $\big(T^{\rm R}+V\big)_-$ is compact.\footnote{In~\cite{DalSorSto-08} it was even shown, using ideas of~\cite{BarFarHelSie-05}, that $\big(T^{\rm R}+V\big)_-$ is Hilbert-Schmidt, but this is not necessary here.} It is indeed a general fact that the quadratic form of a bounded-below operator $A$ is weakly lower semi-continuous for the associated topology, if and only if the essential spectrum of $A$ does not go below 0. One can verify that~\eqref{eq:wlsc} is sufficient to adapt the proof of~\cite{Lewin-04a} to the pseudo-relativistic case.

In the paper of Argaez-Melgaard, the property~\eqref{eq:wlsc} is somewhat proved in the appendix. It is formulated in terms of density matrices, which is fine but not necessary. Even though the same result in the Enstedt-Melgaard paper~\cite[p. 14]{EnsMel-09} simply referred to~\cite{DalSorSto-08}, it is detailed again in the Argaez-Melgaard paper, without even mentioning~\cite{DalSorSto-08}.

What is however \emph{completely wrong} in the Argaez-Melgaard paper is the treatment of the two-body repulsion between the electrons. 
This term is non-negative and the proof of~\cite{Lewin-04a} applies without any difficulty. Instead, Argaez and Melgaard  ``switch to the density operator formulation'' on page 396, and they associate to any MC wave-function $\Psi$ of the form~\eqref{eq:form-MC} the `density' operator
$$\cD=\sum_{1\leq i_1<\cdots <i_N\leq K}a_{i_1...i_N}\,\Big(|\phi_{i_1}\rangle\langle\phi_{i_1}|+\cdots +|\phi_{i_N}\rangle\langle\phi_{i_N}|\Big).$$
Then, on pages 396-397 and in the appendix, they use that the MC total energy is equal to the HF energy of $\cD$. This is \emph{obviously wrong} (the energy is quadratic in the $a_{i_1...i_N}$ but the one-body term in the HF energy is linear). The operator $\cD$ is not even related to the true density matrix $\gamma_\Psi$ (except when all the $a_{i_1...i_N}$ vanish but for one, which equals 1). 

Even if we forget for a moment that the energy cannot be expressed in terms of $\cD$, most of the mathematical arguments of Argaez-Melgaard based on $\cD$ are incorrect. The main difficulty that $\cD$ is not a non-negative operator (the $a_{i_1...i_N}$ can be negative), is avoided in a very questionable fashion. In several places, the authors seem to forget that the $a_{i_1...i_N}$'s have no sign and, when necessary, they suddenly replace $a_{i_1...i_N}$ by the absolute value $|a_{i_1...i_N}|$ (see, e.g., the liminf argument at the bottom of page 402). Even the nonlinear HF interaction term is not weakly lower semi-continuous when the density matrix has no sign, and (A.14) is not correct.

Let me end this Letter by mentioning another issue with the paper of Argaez and Melgaard. Their main result (Proposition 9.1) contains the statement that the limit of a Palais-Smale sequences for the MC energy always has a rank equal to $K-1$ or $K$. The non-relativistic equivalent to this statement was proved independently in~\cite{Friesecke-03b,Lewin-PhD}, based on a previous method of Le Bris~\cite{LeBris-94}. It relies on the fact that the orbitals $\phi_i$'s of a solution to the MC equations are real-analytic away from the nuclear positions. This fact is not proved by Argaez and Melgaard, who instead refer to `a future work' on page 397. So the corresponding statement 5 in Proposition 9.1 is in fact not proved in the paper. Fortunately, the real-analyticity was recently proved\footnote{The proof of this in the Hartree-Fock case was made available on arXiv in the first version of~\cite{DalFouSorSto-11}, before the submission of the paper of Argaez-Melgaard. The observation that the proof works the same 
in the MC case was added in the second version of~\cite{DalFouSorSto-11}, after the publication of~\cite{ArgMel-12}.} by Dall'Acqua, Fournais, {\O}stergaard S{\o}rensen and Stockmeyer in~\cite{DalFouSorSto-11}, and Proposition 9.1 is finally correct.

As a conclusion, the paper of Argaez and Melgaard refers improperly to the literature and some parts of the proofs are wrong.

%%%%%%%%%%%%%%%%%%%%%%%%%%%%%%
%%%%%%%%%%%%%%%%%%%%%%%%%%%%%%
%\bibliographystyle{siam}
%\bibliography{biblio}

\begin{thebibliography}{10}

\bibitem{ArgMel-12}
{\sc C.~Argaez and M.~Melgaard}, {\em Solutions to quasi-relativistic
  multi-configurative {H}artree-{F}ock equations in quantum chemistry},
  Nonlinear Analysis: Theory, Methods \& Applications, 75 (2012), pp.~384--404.

\bibitem{Bach-92}
{\sc V.~Bach}, {\em Error bound for the {H}artree-{F}ock energy of atoms and
  molecules}, Commun. Math. Phys., 147 (1992), pp.~527--548.

\bibitem{BarFarHelSie-05}
{\sc J.~M. Barbaroux, W.~Farkas, B.~Helffer, and H.~Siedentop}, {\em On the
  {H}artree-{F}ock equations of the electron-positron field}, Commun. Math.
  Phys., 255 (2005), pp.~131--159.

\bibitem{CanDefKutLeBMad-03}
{\sc {\'E}.~Canc{\`e}s, M.~Defranceschi, W.~Kutzelnigg, C.~{Le Bris}, and
  Y.~Maday}, {\em Computational quantum chemistry: a primer}, in Handbook of
  numerical analysis, Vol. X, Handb. Numer. Anal., X, North-Holland, Amsterdam,
  2003, pp.~3--270.

\bibitem{DalFouSorSto-11}
{\sc A.~{Dall'Acqua}, S.~{Fournais}, T.~{{\O}stergaard S{\o}rensen}, and
  E.~{Stockmeyer}}, {\em {Real analyticity away from the nucleus of
  pseudorelativistic {H}artree-{F}ock orbitals}}, Analysis \& PDE, to appear
  (2011).

\bibitem{DalSorSto-08}
{\sc A.~{Dall'Acqua}, T.~{{\O}stergaard S{\o}rensen}, and E.~Stockmeyer}, {\em
  {H}artree-{F}ock theory for pseudorelativistic atoms}, Annales Henri
  Poincar\'e, 9 (2008), pp.~711--742.

\bibitem{EnsMel-09}
{\sc M.~Enstedt and M.~Melgaard}, {\em Existence of infinitely many distinct
  solutions to the quasirelativistic {H}artree-{F}ock equations}, Int. J. Math.
  Math. Sci.,  (2009), p.~651871.

\bibitem{FanGho-94}
{\sc G.~Fang and N.~Ghoussoub}, {\em Morse-type information on {P}alais-{S}male
  sequences obtained by min-max principles}, Comm. Pure Appl. Math., 47 (1994),
  pp.~1595--1653.

\bibitem{Friesecke-03}
{\sc G.~Friesecke}, {\em The multiconfiguration equations for atoms and
  molecules: charge quantization and existence of solutions}, Arch. Ration.
  Mech. Anal., 169 (2003), pp.~35--71.

\bibitem{Friesecke-03b}
\leavevmode\vrule height 2pt depth -1.6pt width 23pt, {\em On the infinitude of
  non-zero eigenvalues of the single-electron density matrix for atoms and
  molecules}, R. Soc. Lond. Proc. Ser. A Math. Phys. Eng. Sci., 459 (2003),
  pp.~47--52.

\bibitem{Ghoussoub-93}
{\sc N.~Ghoussoub}, {\em Duality and perturbation methods in critical point
  theory}, vol.~107 of Cambridge Tracts in Mathematics, Cambridge University
  Press, Cambridge, 1993.
\newblock With appendices by David Robinson.

\bibitem{LeBris-94}
{\sc C.~{Le Bris}}, {\em A general approach for multiconfiguration methods in
  quantum molecular chemistry}, Ann. Inst. H. Poincar{\'e} Anal. Non
  Lin{\'e}aire, 11 (1994), pp.~441--484.

\bibitem{Lewin-04a}
{\sc M.~Lewin}, {\em Solutions of the multiconfiguration equations in quantum
  chemistry}, Arch. Ration. Mech. Anal., 171 (2004), pp.~83--114.

\bibitem{Lewin-PhD}
\leavevmode\vrule height 2pt depth -1.6pt width 23pt, {\em Some nonlinear
  models in Quantum Mechanics}, PhD thesis, University of Paris-Dauphine, June
  2004.

\bibitem{LewSieVug-97}
{\sc R.~T. Lewis, H.~Siedentop, and S.~Vugalter}, {\em The essential spectrum
  of relativistic multi-particle operators}, Ann. Inst. Henri Poincar{\'e}, 67
  (1997), pp.~1--28.

\bibitem{Lieb-81}
{\sc E.~H. Lieb}, {\em Variational principle for many-fermion systems}, Phys.
  Rev. Lett., 46 (1981), pp.~457--459.

\bibitem{LieSim-77}
{\sc E.~H. Lieb and B.~Simon}, {\em The {H}artree-{F}ock theory for {C}oulomb
  systems}, Commun. Math. Phys., 53 (1977), pp.~185--194.

\bibitem{LieThi-84}
{\sc E.~H. Lieb and W.~E. Thirring}, {\em Gravitational collapse in quantum
  mechanics with relativistic kinetic energy}, Ann. Physics, 155 (1984),
  pp.~494--512.

\bibitem{LieYau-87}
{\sc E.~H. Lieb and H.-T. Yau}, {\em The {C}handrasekhar theory of stellar
  collapse as the limit of quantum mechanics}, Commun. Math. Phys., 112 (1987),
  pp.~147--174.

\bibitem{LieYau-88}
\leavevmode\vrule height 2pt depth -1.6pt width 23pt, {\em The stability and
  instability of relativistic matter}, Commun. Math. Phys., 118 (1988),
  pp.~177--213.

\bibitem{Lions-87}
{\sc P.-L. Lions}, {\em Solutions of {H}artree-{F}ock equations for {C}oulomb
  systems}, Commun. Math. Phys., 109 (1987), pp.~33--97.

\bibitem{Solovej-91}
{\sc J.~P. Solovej}, {\em {Proof of the ionization conjecture in a reduced
  Hartree-Fock model.}}, Invent. Math., 104 (1991), pp.~291--311.

\bibitem{Zhislin-06}
{\sc G.~M. Zhislin}, {\em The spectrum of {H}amiltonians of pseudorelativistic
  electrons of molecules in spaces of functions with permutational and point
  symmetry}, Funktsional. Anal. i Prilozhen., 40 (2006), pp.~65--69.

\bibitem{ZhiVug-02}
{\sc G.~M. Zhislin and S.~A. Vugalter}, {\em On the discrete spectrum of
  {H}amiltonians for pseudorelativistic electrons}, Izv. Ross. Akad. Nauk Ser.
  Mat., 66 (2002), pp.~71--102.

\end{thebibliography}

\end{document}